\documentclass[epj]{svjour}

\usepackage{graphics}

\usepackage{amssymb}

\begin{document}

\title{Two neutrino double-$\beta $ decay of $94\leq A\leq 150$ nuclei for
the 0$^{+}\rightarrow $2$^{+}$ transition}

\author{Yash Kaur Singh\inst{1} \and 
R. Chandra\inst{1} \and 
\thanks{\emph{Corresponding Author:} ramesh.luphy@gmail.com}
P.K. Raina\inst{2} \and 
P.K. Rath\inst{3}
}

\institute {Department of Applied Physics, Babasaheb Bhimrao Ambedkar 
University, Lucknow, India. \and
Department of Physics, Indian Institute of Technology, Ropar, Rupnagar
- 140001, India \and
Department of Physics, University of Lucknow, Lucknow-226007, India.}

\date{Received: date / Revised version: date}

\abstract{
Within the PHFB approach, the $0^{+}\rightarrow 2^{+}$ transition of two
neutrino double-$\beta $ decay of $^{94,96}$Zr, $^{100}$Mo, $^{104}$Ru, 
$^{110}$Pd, $^{128,130}$Te and $^{150}$Nd isotopes is studied employing wave
functions generated with four different parametrizations of the pairing plus
multipole type of two-nucleon interaction and the summation method. In
comparison to the $0^{+}\rightarrow 0^{+}$ transition, the nuclear
transition matrix elements $M_{2\nu }(2^{+})$ are quite sensitive to the
deformation of the yrast 2$^{+}$ state. Consideration of the available
theoretical and experimental results suggest that the observation of the 0$%
^{+}\rightarrow $2$^{+}$ transition of $2\nu \beta ^{-}\beta ^{-}$ decay may
be possible in $^{96}$Zr, $^{100}$Mo, $^{130}$Te and $^{150}$Nd isotopes. 
The effect of deformation on the $M_{2\nu }(2^{+})$ is also studied.
     }

\maketitle

\section{Introduction}

The nuclear double beta $(\beta \beta )$ decay is a convenient tool to test
the validity of models employed in the nuclear structure studies and probe
the physics beyond standard model of electroweak unification (SM). Over the
past years, the theoretical as well as experimental studies of the $\beta
\beta $ decay has attracted a lot of attention and excellently reviewed in Ref. \cite%
{verg16,saak13,henn16,ostr16,enge17} and references therein. The two
neutrino double beta $(2\nu \beta \beta )$ decay is a second order process
in weak interaction and is allowed in the SM. The neutrinoless double beta $%
(0\nu \beta \beta )$ decay is far more interesting as it involves the
Majorana neutrinos and violation of the lepton number conservation by two
units. The observation of the $0\nu \beta \beta $ decay can not only
establish the Majorana nature of neutrinos but also provide information on
the physics beyond the SM \cite{klap06}.

The $\beta \beta $ decay can occur in four different modes namely,
double-electron $(\beta ^{-}\beta ^{-})$ emission , double -positron $(\beta
^{+}\beta ^{+})$ emission, electron positron conversion $(\varepsilon \beta
^{+})$ and double-electron capture $(\varepsilon \varepsilon )$. The latter
three processes are energetically competing. In the allowed approximation,
the $0^{+}\rightarrow 1^{+}$ transition is much less probable than the $%
0^{+}\rightarrow 0^{+}$ and $0^{+}\rightarrow 2^{+}$ transitions. The
observation of $0\nu \beta \beta $ decay for the $0^{+}\rightarrow 2^{+}$
transition can distinguish between the mechanisms involving the mass of the
Majorana neutrinos and the right handed currents \cite{tomo91}. The
theoretical implications and experimental aspects of the ground to the excited 2$^{+}$ state transition of the $\beta \beta $ decay have been
excellently reviewed over past years \cite{suho98}.

Out of 35 possible candidates, the $0^{+}\rightarrow 0^{+}$transition of $%
2\nu \beta ^{-}\beta ^{-}$ decay has been observed for twelve nuclei \cite%
{saak13,bara15} and limits on the half-lives $T_{1/2}^{2\nu }$ of a number of
isotopes for the $0^{+}\rightarrow 0^{+}$ and $0^{+}\rightarrow 2^{+}$
transitions have already been given \cite{tret02}. The inverse half- life of 
$2\nu \beta ^{-}\beta ^{-}$ decay is a product of the phase space factor and
model dependent nuclear transition matrix elements (NTMEs) $M_{2\nu }$. The
phase space factors have been calculated employing the exact Dirac wave
functions in conjunction with finite nuclear size and screening effects \cite%
{koti12,stoi13}. Using the observed experimental half-lives for the $%
0^{+}\rightarrow 0^{+}$ transition, the NTMEs $M_{2\nu }$ has been extracted 
\cite{bara15} and in all cases of $2\nu \beta ^{-}\beta ^{-}$ decay, it has
been observed that the NTMEs $M_{2\nu }(0^{+})$ are sufficiently quenched \cite{bare15}.
The main motive of all theoretical calculations is to understand the
physical mechanism responsible for the observed suppression of $M_{2\nu
}(0^{+})$. Hence, the validity of different nuclear models can
be tested by calculating $M_{2\nu }(0^{+})$ and comparing them with the
experimental value.

The $0^{+}\rightarrow 2^{+}$ transition of $2\nu \beta ^{-}\beta ^{-}$ decay
has not been experimentally observed so far. The marked variation in the
theoretically calculated NTMEs $M_{2\nu }$($2^{+}$) for the $0^{+}\rightarrow 2^{+}$
transition using different nuclear models is a general feature 
\cite{suho98}. For example, the available results for $M_{2\nu }(2^{+})$ of $%
^{96}$Zr show that the calculated NTMEs within QRPA \cite%
{bara96,radu07,unlu14}, RQRPA(WS) \cite{toiv97}, RQRPA (AWS) \cite{toiv97},
and SRPA(WS) \cite{stoi96}, differ by a factor of 341. Hence, the
observation of the $0^{+}\rightarrow 2^{+}$ transition of $2\nu \beta
^{-}\beta ^{-}$ decay can constrain the validity of different nuclear models
employed in the calculation of NTMEs. Alternatively, a reliable theoretical
prediction will supplement the experimental designing and planning to study
this particular mode of $2\nu \beta ^{-}\beta ^{-}$ decay.

Employing the pnQRPA model, it has been shown by Raduta $et$ $al$. \cite%
{radu07} that the inclusion of deformation in the mean field can reduce the
NTMEs $M_{2\nu }(2^{+})$ up to a factor of 341. In the PHFB model, the
pairing and deformation degrees of freedom are treated simultaneously on
equal footing. However, the structure of intermediate odd-odd nuclei can not
be studied in the present version of the PHFB model. In spite of this
limitation, the PHFB model has been successfully applied to study the $%
0^{+}\rightarrow 0^{+}$ transition of $2\nu \beta ^{-}\beta ^{-}$ decay \cite%
{chan05,sing07} in conjunction with the summation method \cite{civi93}. This
has motivated us to apply the same set of wave functions to study the $%
0^{+}\rightarrow 2^{+}$ transition of $2\nu \beta ^{-}\beta ^{-}$ decay of $%
^{94,96}$Zr, $^{100}$Mo, $^{104}$Ru, $^{110}$Pd, $^{128,130}$Te and $^{150}$%
Nd isotopes in the mass range $90\leq A\leq 150$.

The theoretical formalism to calculate the half-life for the 0$%
^{+}\rightarrow $ 2$^{+}$ transition of $2\nu \beta ^{-}\beta ^{-}$ decay $%
T_{1/2}^{2\nu }(0^{+}\rightarrow 2^{+})$ in 2n mechanism has been given in
Refs. \cite{tomo91,haxt84,doi85}. Using the summation method \cite{civi93},
the $0^{+}\rightarrow 0^{+}$ and $0^{+}\rightarrow 2^{+}$ transitions of 2$%
\nu \beta ^{-}\beta ^{-}$ mode has already been studied by Hirsch $et$ $al$.
in the pseudo-SU(3) model \cite{hirs95,hirs95b}. Presently, the summation
method applied to the study of $0^{+}\rightarrow 0^{+}$ transition of 2$\nu
\beta ^{-}\beta ^{-}$ decay within the PHFB model \cite{chan05,sing07} has
been extended to the $0^{+}\rightarrow 2^{+}$ transition. In sect. 2, we
outline the theoretical formalism to calculate the half life $T_{1/2}^{2\nu
}(2^{+})$ of 2$\nu \beta ^{-}\beta ^{-}$ decay. The results are presented
and discussed in sect. 3. The final conclusions are given in sect. 4.

\section{Theoretical Framework}

The half life for the $0^{+}\rightarrow 2^{+}$ transition of 2$\nu \beta
^{-}\beta ^{-}$ decay $T_{1/2}^{2\nu }(2^{+})$ in 2n mechanism is given by%
\begin{equation}
\left[ T_{1/2}^{2\nu }(2^{+})\right] ^{-1}=G_{2\nu }(2^{+})\left\vert
M_{2\nu }(2^{+})\right\vert ^{2}
\end{equation}%
where the integrated kinematical factor $G_{2\nu }(2^{+})$ has been
calculated with good accuracy \cite{paho13}. The model dependent NTME $%
M_{2\nu }(2^{+})$ is given by%
\begin{equation}
M_{2\nu }(2^{+})=\sqrt{\frac{1}{3}}\sum\limits_{N}\frac{\left\langle
2^{+}\left\Vert \sigma \tau ^{+}\right\Vert 1_{N}^{+}\right\rangle
\left\langle 1_{N}^{+}\left\Vert \sigma \tau ^{+}\right\Vert
0^{+}\right\rangle }{\left[ E_{0}+E_{N}-E_{I}\right] ^{3}}
\end{equation}%
where

\begin{equation}
E_{0}=\frac{1}{2}(E_{I}-E_{F})=\frac{1}{2}Q_{\beta \beta }+m_{e})
\end{equation}%
Presently, the summation over the intermediate states is performed using the
summation method \cite{civi93}. Extending the summation method already
applied to the $0^{+}\rightarrow 0^{+}$ transition of $2\nu \beta ^{-}\beta
^{-}$ decay \cite{chan05,sing07} to the $0^{+}\rightarrow 2^{+}$ transition,
the NTME $M_{2\nu }(2^{+})$ is written as%
\begin{equation}
M_{2\nu }(2^{+})=\sqrt{5}\sum\limits_{\pi ,\nu }\frac{\left\langle
2_{F}^{+}\left\Vert [\mathbf{\sigma }\otimes \mathbf{\sigma ]}^{(2)}\tau
^{+}\tau ^{+}\right\Vert 0_{I}^{+}\right\rangle }{\left[ E_{0}+\varepsilon
(n_{\pi },l_{\pi },j_{\pi })-\varepsilon (n_{\nu },l_{\nu },j_{\nu })\right]
^{3}}
\end{equation}%
and this expression is same as Hirsch $et$ $al$. \cite{hirs95b}.

As each proton-neutron excitation is considered according to its spin-flip
or non--spin-flip character, the use of the summation method in the present
context goes beyond the closure approximation. The spin-orbit splitting is
explicitly included in the energy denominator, and hence, the PHFB formalism
in conjunction with the summation method goes beyond that previously
employed in the pseudo SU(3) model \cite{hirs95,hirs95b}. In the PHFB model,
the NTME $M_{2\nu }(2^{+})$ for the 0$^{+}\rightarrow 2^{+}$ transition of $%
2\nu \beta ^{-}\beta ^{-}$ decay is calculated using

\begin{eqnarray}
M_{2\nu } &=&\sum\limits_{\pi ,\nu }\frac{\langle {\Psi _{00}^{J_{f}=2}}||[%
\mathbf{\sigma }\otimes \mathbf{\sigma ]}^{(2)}\tau ^{+}\tau ^{+}||{\Psi
_{00}^{J_{i}=0}}\rangle }{[E_{0}+\varepsilon (n_{\pi },l_{\pi },j_{\pi
})-\varepsilon (n_{\nu },l_{\nu },j_{\nu })]^{3}}  \nonumber \\
&=&\left[ n_{(Z,N)}^{Ji=2}n_{(Z+2,N-2)}^{J_{f}=0}\right] ^{-1/2}\int%
\limits_{0}^{\pi }n_{(Z,N),(Z+2,N-2)}(\theta )  \nonumber \\
&&\times \sum_{\mu }\left[ 
\begin{array}{lll}
J_{i} & 2 & J_{f} \\ 
-\mu & \mu & 0%
\end{array}%
\right] d_{\mu 0}^{J_{i}}\left( \theta \right)  \nonumber \\
&&\times \sum\limits_{\alpha \beta \gamma \delta }\frac{\left\langle \alpha
\beta \left\vert \lbrack \mathbf{\sigma }\otimes \mathbf{\sigma ]}^{(2)}\tau
^{+}\tau ^{+}\right\vert \gamma \delta \right\rangle }{[E_{0}+\varepsilon
_{\alpha }(n_{\pi },l_{\pi },j_{\pi })-\varepsilon _{\gamma }(n_{\nu
},l_{\nu },j_{\nu })]^{3}}  \nonumber \\
&&\times \sum\limits_{\varepsilon \eta }\left[ \left( 1+F_{Z,N}^{(\pi
)}(\theta )f_{Z+2,N-2}^{(\pi )\ast }\right) \right] _{\varepsilon \alpha
}^{-1}\left( f_{Z+2,N-2}^{(\pi )\ast }\right) _{\varepsilon \beta } 
\nonumber \\
&&\times \left[ \left( 1+F_{Z,N}^{(\nu )}(\theta )f_{Z+2,N-2}^{(\nu )\ast
}\right) \right] _{\gamma \eta }^{-1}\left( F_{Z,N}^{(\nu )\ast }\right)
_{\eta \delta }sin\theta d\theta  \label{m2n}
\end{eqnarray}%
and the expressions for $n^{J}$, $n_{(Z,N),(Z+2,N-2)}{(\theta )}$, $f_{Z,N}$
and $F_{Z,N}(\theta )$ are given in Ref. \cite{chan05,sing07}.

\section{Results and discussions}

The model space, single particle energies (SPE's), parameters of pairing
plus multipolar type of effective two-body interaction have already been
discussed in Refs. \cite{chan05,sing07,chan09,rath10}. Specifically, the
effective Hamiltonian is written as \cite{chan09} 
\begin{equation}
H=H_{sp}+V(P)+\zeta_{qq}\left[V(QQ)+V(HH)\right],
\end{equation}%
where $H_{sp}$, $V(P)$, $V(QQ)$ and $V(HH)$ denote the single particle
Hamiltonian, the pairing, quadrupole-quadrupole and
hexadecapole-hexadecapole parts of the effective two-body interaction,
respectively. The $\zeta_{qq}$ is an arbitrary parameter and the final 
results are obtained by setting the $\zeta_{qq}=1$. The purpose of 
introducing $\zeta_{qq}$ is to study the role of deformation by varying 
the strength of the {\it QQ} and {\it HH} interactions. In the $QQ$ part 
of the effective two-body interaction $V(QQ)$,
the strengths of the proton-proton, the neutron-neutron and the
proton-neutron interactions are denoted by $\chi _{2pp},\chi _{2nn}$ and $%
\chi _{2pn}$, respectively. By reproducing the experimental excitation
energies $E_{2^{+}}$ of the $\ $2$^{+}$ state in two alternative ways
provides two different parametrization of the $QQ$ interaction, namely $PQQ1$
\cite{chan05,sing07} and $PQQ2$ \cite{rath10}. The inclusion of the
hexadecapolar $HH$ part of the effective interaction adds two additional
parametrizations, namely $PQQHH1$ \cite{chan09} and $PQQHH2$ \cite{rath10}.

In Ref. \cite{chan05,sing07,chan09,rath10}, the reliability of wave
functions generated with four different parametrizations of the effective
two-body interaction, namely \textit{PQQ}1, \textit{PQQHH}1, \textit{PQQ}2
and \textit{PQQHH}2 was tested by comparing the theoretically calculated
results for a number of spectroscopic properties, namely the yrast spectra,
reduced $B(E2$:$0^{+}\rightarrow 2^{+})$ transition probabilities,
quadrupole moments $Q(2^{+})$ and $g$-factors $g(2^{+})$ of $^{94,96}$Zr, $%
^{94,96,100}$Mo, $^{100,104}$Ru, $^{104,110}$Pd, $^{110}$Cd, $^{128,130}$Te, 
$^{128,130}$Xe, $^{150}$Nd and $^{150}$Sm isotopes with the available
experimental data. In addition, the calculated $M_{2\nu }$ and corresponding 
$T_{1/2}^{2\nu }$ for the $0^{+}\rightarrow 0^{+}$ transition were compared
with the available experimentally observed results. Presently, the same set
of wave functions are employed to calculate the NTMEs $M_{2\nu }(2^{+})$.

In table 1, the NTMEs $M_{2\nu }(2^{+})$ calculated with wave functions
generated with four different parametrizations of effective two-body
interactions are presented. Although, there are only a set of four NTMEs $%
M_{2\nu }(2^{+})$ for a statistical analysis, the estimated average NTMEs $%
\overline{M}_{2\nu }(2^{+})$ and uncertainties $\Delta \overline{M}_{2\nu
}(2^{+})$ are given in the same table 1. The maximum uncertainty $\Delta 
\overline{M}_{2\nu }(2^{+})$ in the average NTMEs $\overline{M}_{2\nu
}(2^{+})$ turns out to be about 45\%, which implies that the NTMEs $M_{2\nu
}(2^{+})$ are highly sensitive to the deformation content of the intrinsic
wave functions. The phase space factors $G_{2\nu }(2^{+})$ have been
calculated by Pahomi $et$ $al$. \cite{paho13} for most of the prospective $%
2\nu \beta ^{-}\beta ^{-}$ emitters. However, the $G_{2\nu }(2^{+})$ of $%
^{94}$Zr and $^{104}$Ru isotopes are not available. We calculate them by
adopting the prescription of Suhonen and Civitarese \cite{suho98} using
axial vector coupling constant $g_{A}=1.2701$ \cite{beri12}. 
The calculated $G_{2\nu }(2^{+})$ for the $0^{+}\rightarrow 2^{+}$ 
transition of $2\nu \beta ^{-}\beta _{{}}^{-}$ decay of $^{94}$Zr and 
$^{104}$Ru are 6.801$\times10^{-30}$ y$^{-1}$ and 
9.625$\times 10^{-25}$ y$^{-1}$, respectively. 
\begin{table*}[htbp]
\caption{Calculated NTMEs $M_{2\nu }(2^{+})$ within the PHFB model and their
average $\overline{M}_{2\nu }(2^{+})$ along with standard deviation $\Delta 
\overline{M}_{2\nu }(2^{+})$.}
\label{tab1}
\begin{tabular}{rcccccccccccc}
\hline\noalign{\smallskip}
{\small Nuclei} &~~~~ & \multicolumn{7}{c}{$M_{2\nu }(2^{+})$} &~~~~~ & 
$\overline{M}_{2\nu }(2^{+})$ &~~~~~ & $\Delta \overline{M}_{2\nu }(2^{+})$ \\ 
\cline{3-9}
&~~~~ & {\small {\it PQQ1}} &~~~ & {\small {\it PQQHH1}} &~~~ & 
{\small {\it PQQ2}} &~~~ & {\small {\it PQQHH2}} &  &  &  &  \\ \hline
$^{94}${\small Zr} &  & {\small 1.44}$\times ${\small 10}$^{-4}$ &  &
{\small 1.08}$\times ${\small 10}$^{-4}$ &  & {\small 4.08}$\times ${\small %
10}$^{-5}$ &  & {\small 1.02}$\times ${\small 10}$^{-4}$ &  & {\small 9.88}$%
\times ${\small 10}$^{-5}$ &  & {\small 4.30}$\times ${\small 10}$^{-5}$ \\
$^{96}${\small Zr} &  & {\small 9.71}$\times ${\small 10}$^{-5}$ &  &
{\small 1.09}$\times ${\small 10}$^{-4}$ &  & {\small 9.15}$\times ${\small %
10}$^{-5}$ &  & {\small 1.02}$\times ${\small 10}$^{-4}$ &  & {\small 9.98}$%
\times ${\small 10}$^{-5}$ &  & {\small 0}.{\small 74}$\times ${\small 10}$%
^{-5}$ \\
$^{100}${\small Mo} &  & {\small 1.95}$\times ${\small 10}$^{-5}$ &  &
{\small 2.52}$\times ${\small 10}$^{-5}$ &  & {\small 2.02}$\times ${\small %
10}$^{-5}$ &  & {\small 1.05}$\times ${\small 10}$^{-5}$ &  & {\small 1.89}$%
\times ${\small 10}$^{-5}$ &  & {\small 0.61}$\times ${\small 10}$^{-5}$ \\
$^{104}${\small Ru} &  & {\small 3.30}$\times ${\small 10}$^{-5}$ &  &
{\small 4.22}$\times ${\small 10}$^{-5}$ &  & {\small 3.05}$\times ${\small %
10}$^{-5}$ &  & {\small 3.96}$\times ${\small 10}$^{-5}$ &  & {\small 3.63}$%
\times ${\small 10}$^{-5}$ &  & {\small 0.55}$\times ${\small 10}$^{-5}$ \\
$^{110}${\small Pd} &  & {\small 1.21}$\times ${\small 10}$^{-4}$ &  &
{\small 1.33}$\times ${\small 10}$^{-4}$ &  & {\small 1.12}$\times ${\small %
10}$^{-4}$ &  & {\small 1.10}$\times ${\small 10}$^{-4}$ &  & {\small 1.19}$%
\times ${\small 10}$^{-4}$ &  & {\small 0.10}$\times ${\small 10}$^{-4}$ \\
$^{128}${\small Te} &  & {\small 1.19}$\times ${\small 10}$^{-6}$ &  &
{\small 2.89}$\times ${\small 10}$^{-6}$ &  & {\small 1.54}$\times ${\small %
10}$^{-6}$ &  & {\small 2.65}$\times ${\small 10}$^{-6}$ &  & {\small 2.07}$%
\times ${\small 10}$^{-6}$ &  & {\small 0.83}$\times ${\small 10}$^{-6}$ \\
$^{130}${\small Te} &  & {\small 7.72}$\times ${\small 10}$^{-7}$ &  &
{\small 1.86}$\times ${\small 10}$^{-6}$ &  & {\small 8.55}$\times ${\small %
10}$^{-7}$ &  & {\small 1.87}$\times ${\small 10}$^{-6}$ &  & {\small 1.34}$%
\times ${\small 10}$^{-6}$ &  & {\small 0.61}$\times ${\small 10}$^{-6}$ \\
$^{150}${\small Nd} &  & {\small 6.32}$\times ${\small 10}$^{-6}$ &  &
{\small 5.84}$\times ${\small 10}$^{-6}$ &  & {\small 5.74}$\times ${\small %
10}$^{-6}$ &  & {\small 5.54}$\times ${\small 10}$^{-6}$ &  & {\small 5.86}$%
\times ${\small 10}$^{-6}$ &  & {\small 0.33}$\times ${\small 10}$^{-6}$ \\
\noalign{\smallskip}\hline
\end{tabular}
\end{table*}

A suppression of NTMEs $M_{2\nu }(0^{+})$ for $2\nu \beta ^{-}\beta ^{-}$
decay with respect to the spherical case has been reported when the parent
and daughter nuclei have different deformations \cite{chan09,alva04,mene09}.
To investigate this effect for the $0^{+}\rightarrow 2^{+}$ transition, we
present the NTMEs $M_{2\nu }(2^{+})$ for the $2\nu \beta ^{-}\beta ^{-}$
decay of $^{94,96}$Zr, $^{100}$Mo, $^{104}$Ru, $^{110}$Pd, $^{128,130}$Te
and $^{150}$Nd isotopes in fig.~1 as a function of the difference in the
deformation parameter $\Delta \beta _{2}=\beta _{2}(parent)-\beta
_{2}(daughter)$ between the parent and daughter nuclei. The NTMEs $M_{2\nu
}(2^{+})$ are calculated by keeping the deformation for parent nuclei fixed
at $\zeta _{qq}=1$ and changing the deformation of daughter nuclei by
varying $\zeta _{qq}$ in the range 0.0 -- 1.5. It can be observed that in all 
cases but for $^{128,130}$Te, the largest NTMEs correspond to the 
$\left\vert \Delta \beta _{2}\right\vert$ close to zero.
With further increase in deformation, the NTMEs decrease with increase in $\left\vert \Delta \beta _{2}\right\vert$.
\begin{figure*}[htbp]
\begin{tabular}{cc}
\resizebox{0.45\textwidth}{!}{\includegraphics{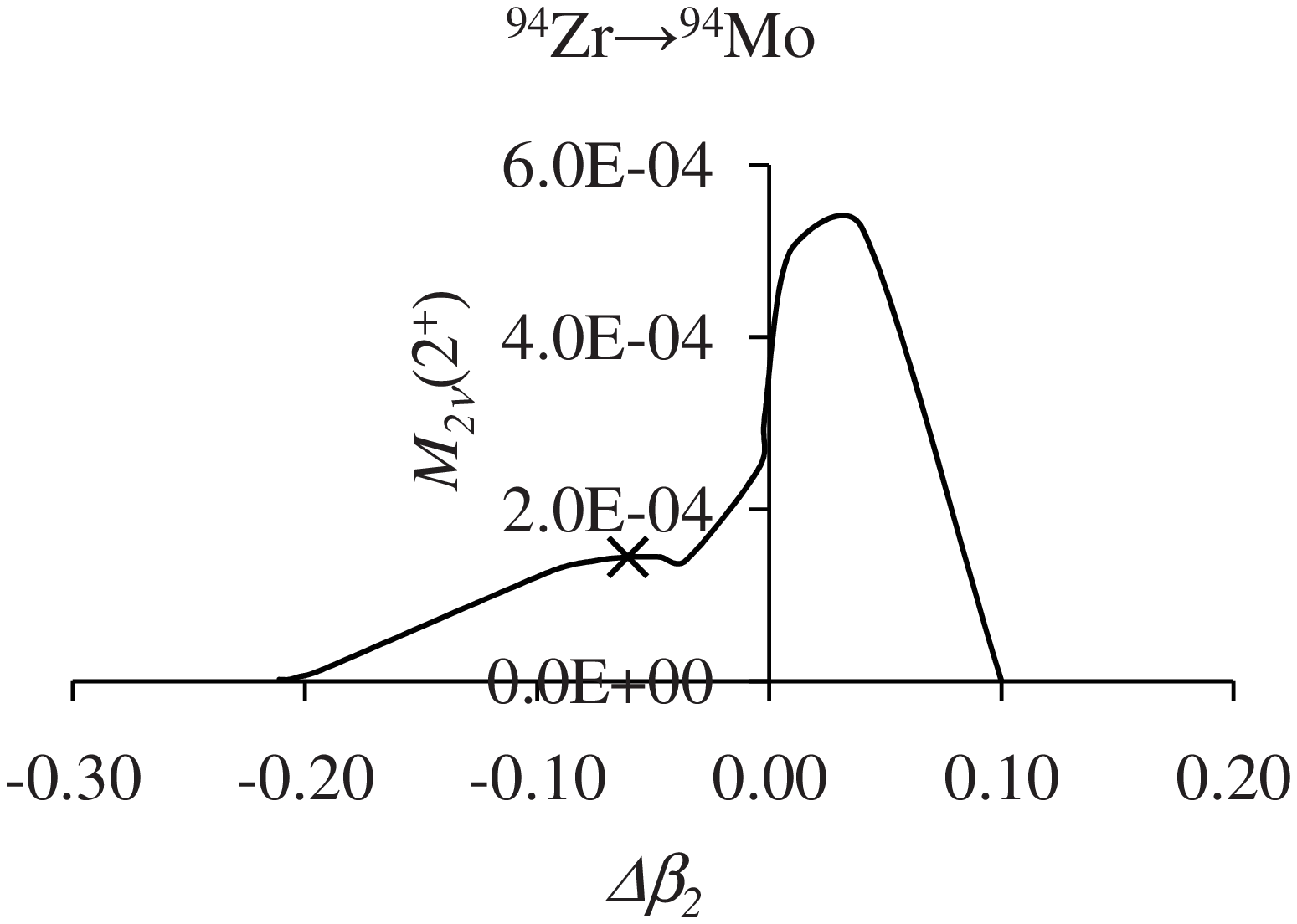}} &
\resizebox{0.45\textwidth}{!}{\includegraphics{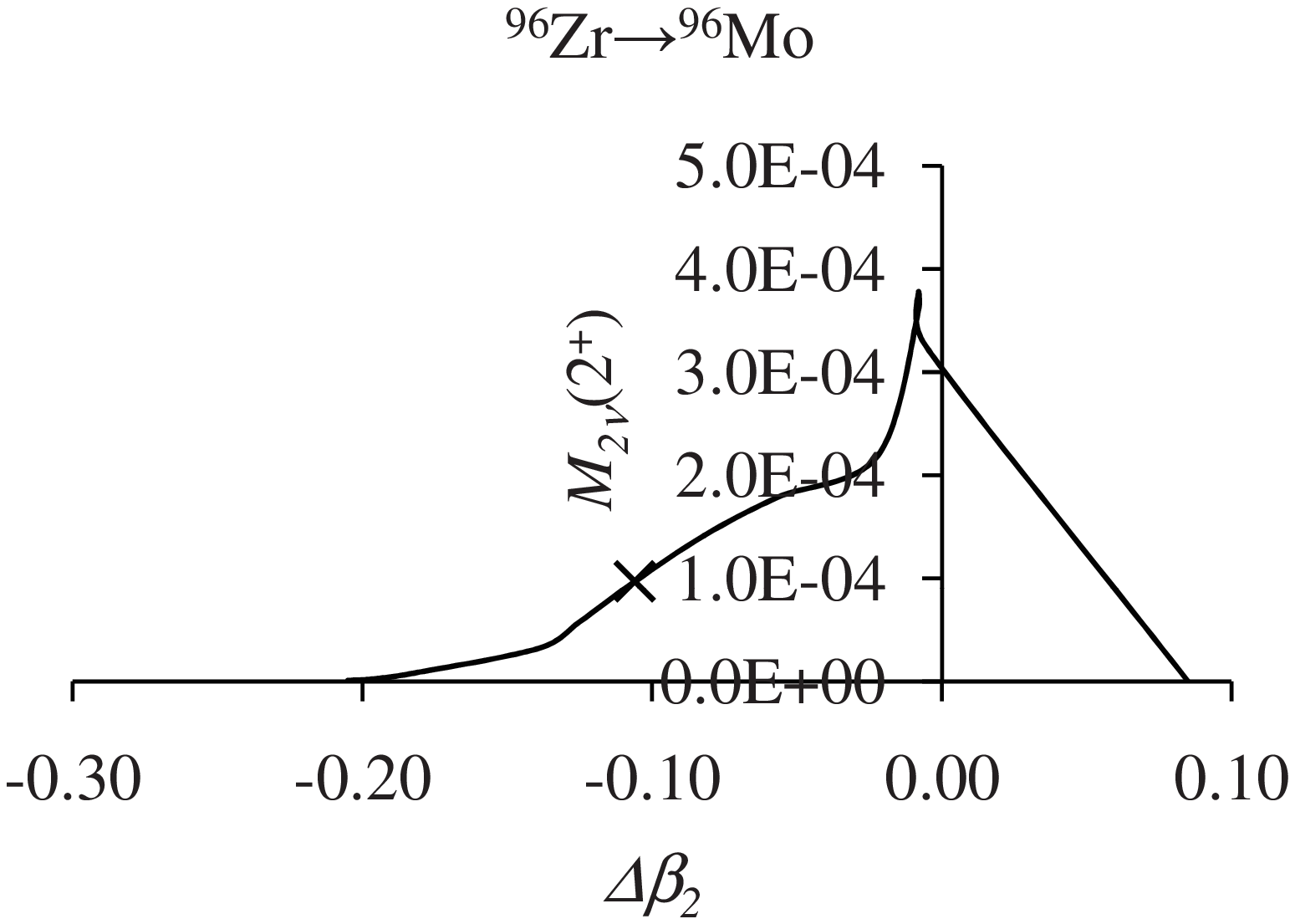}} \\
\resizebox{0.45\textwidth}{!}{\includegraphics{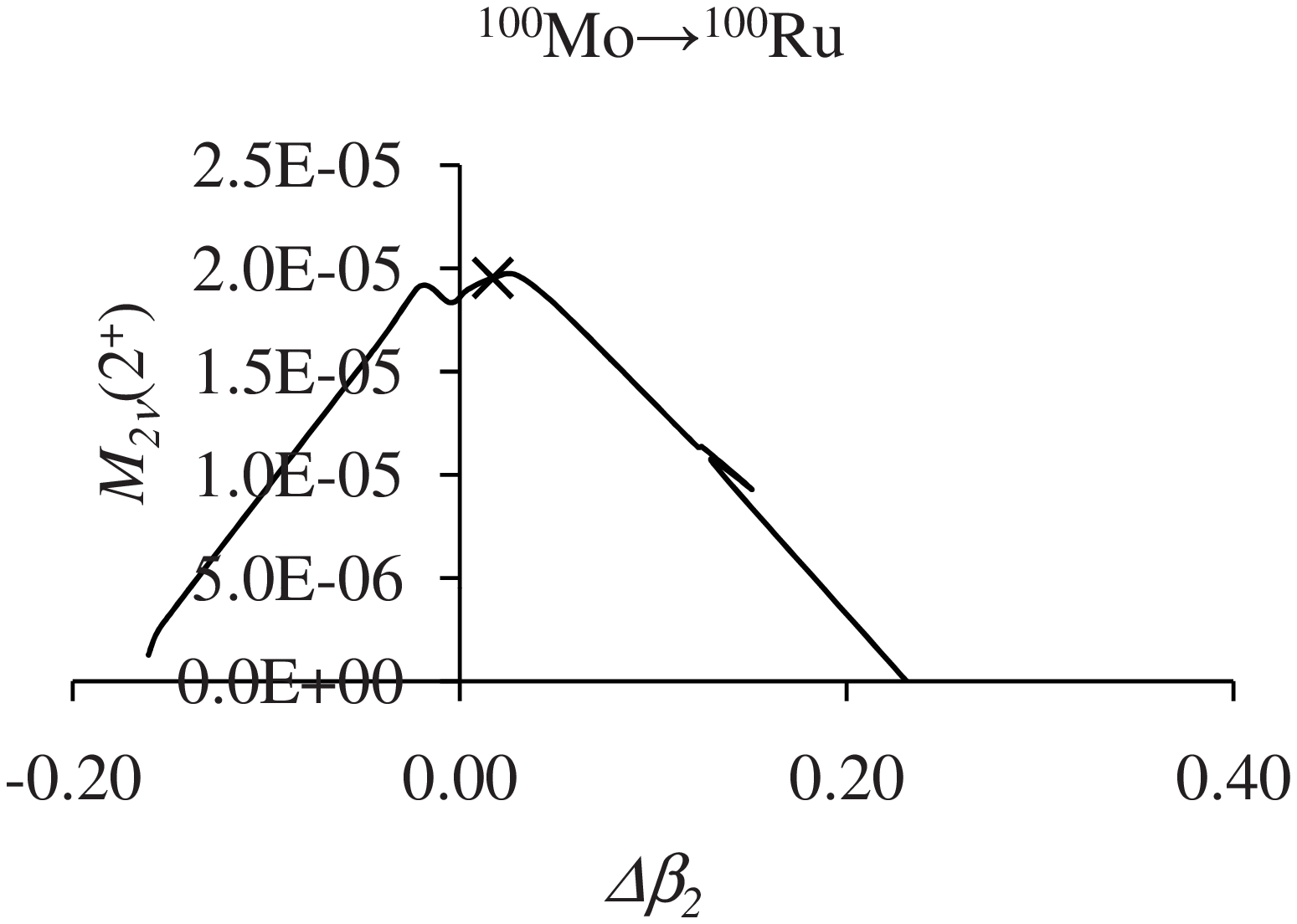}} &
\resizebox{0.45\textwidth}{!}{\includegraphics{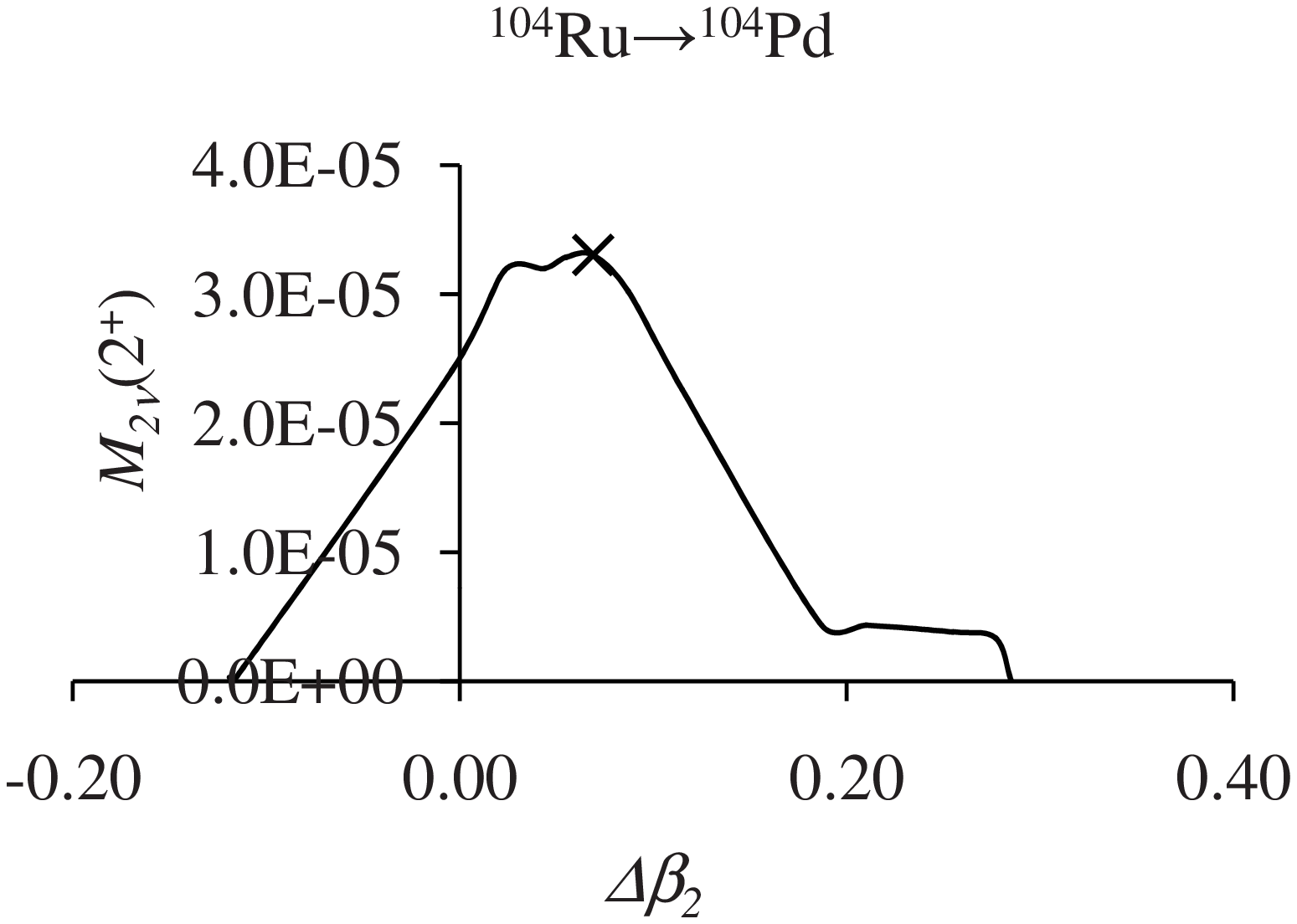}} \\
\resizebox{0.45\textwidth}{!}{\includegraphics{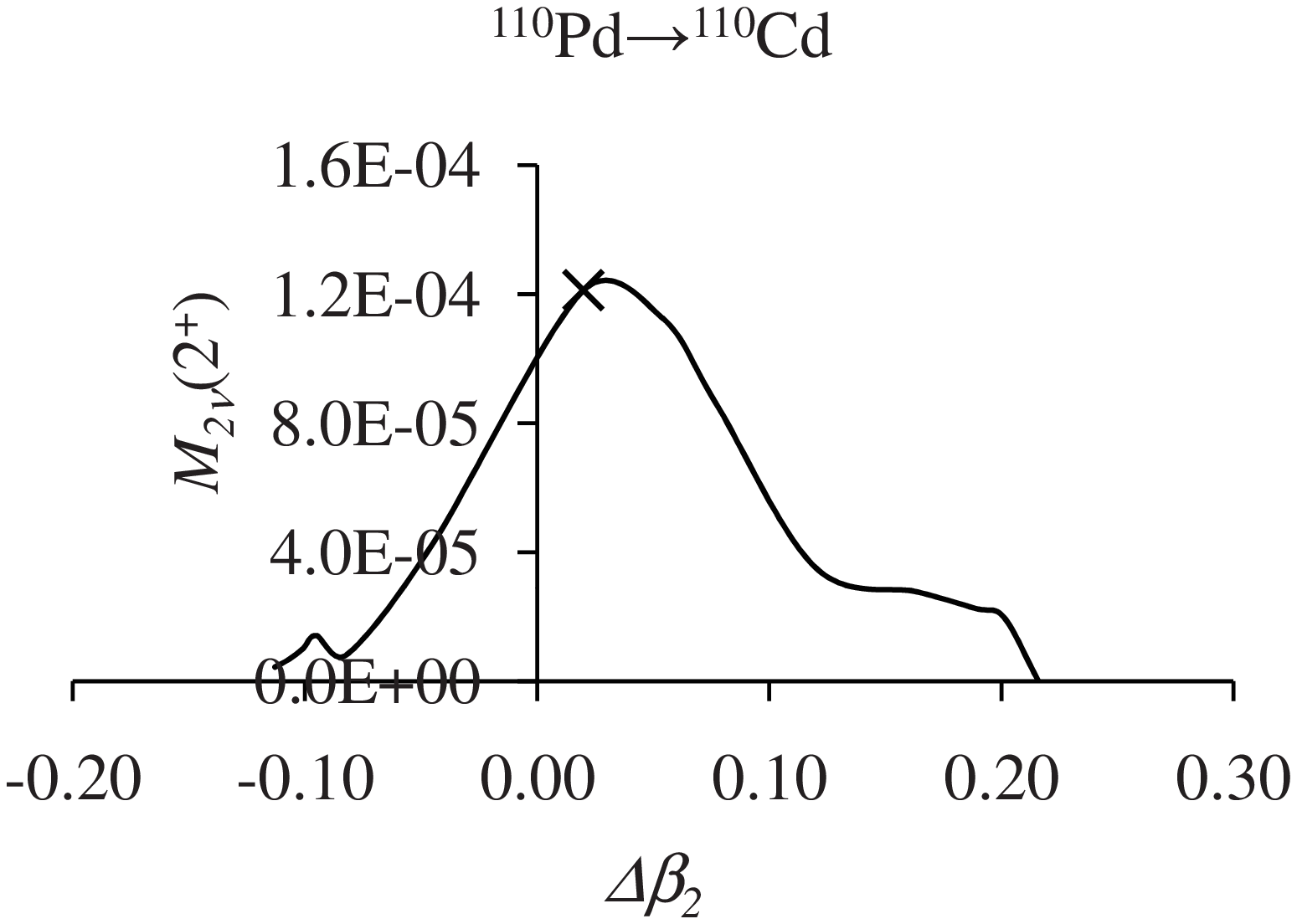}} &
\resizebox{0.45\textwidth}{!}{\includegraphics{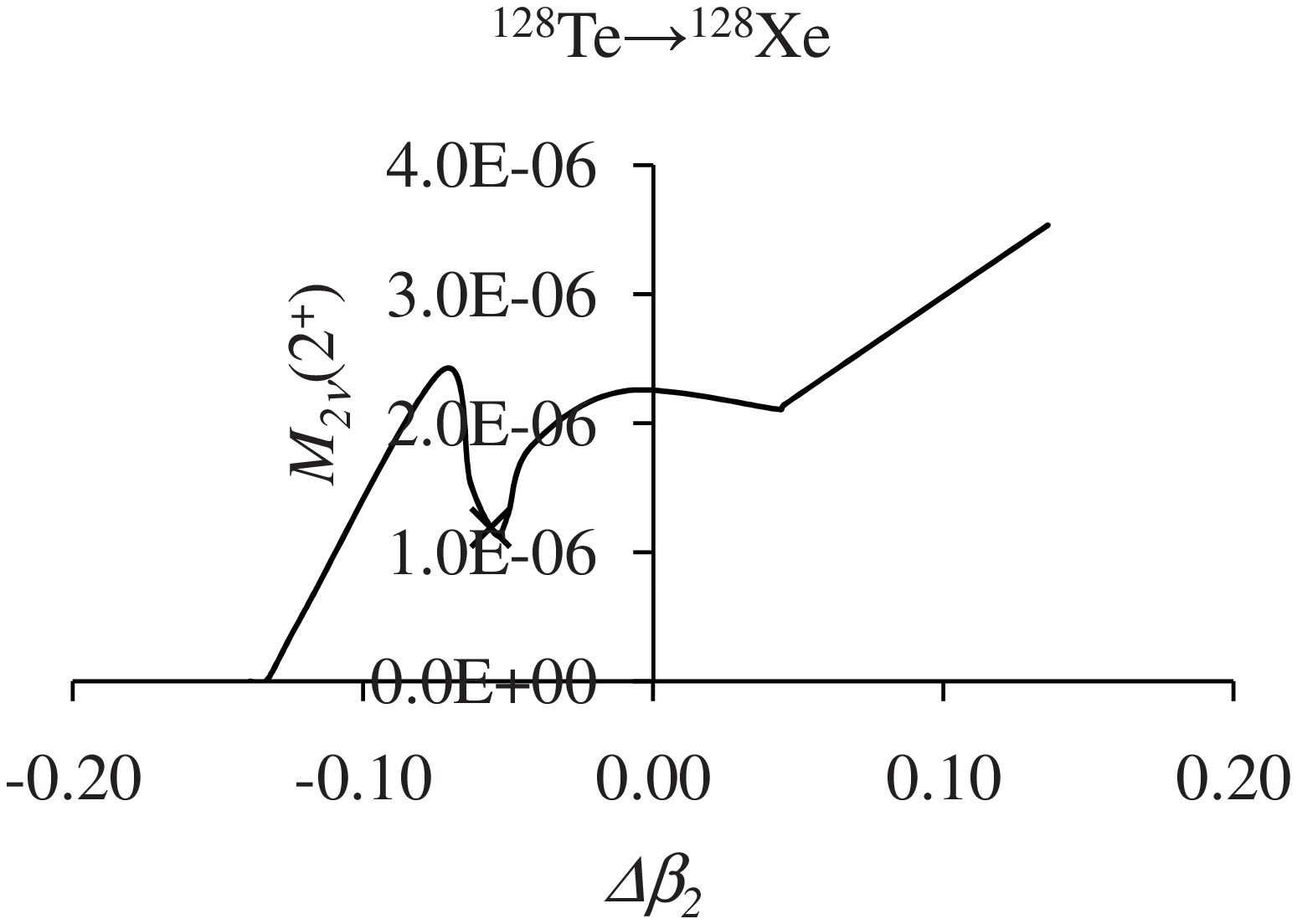}} \\
\resizebox{0.45\textwidth}{!}{\includegraphics{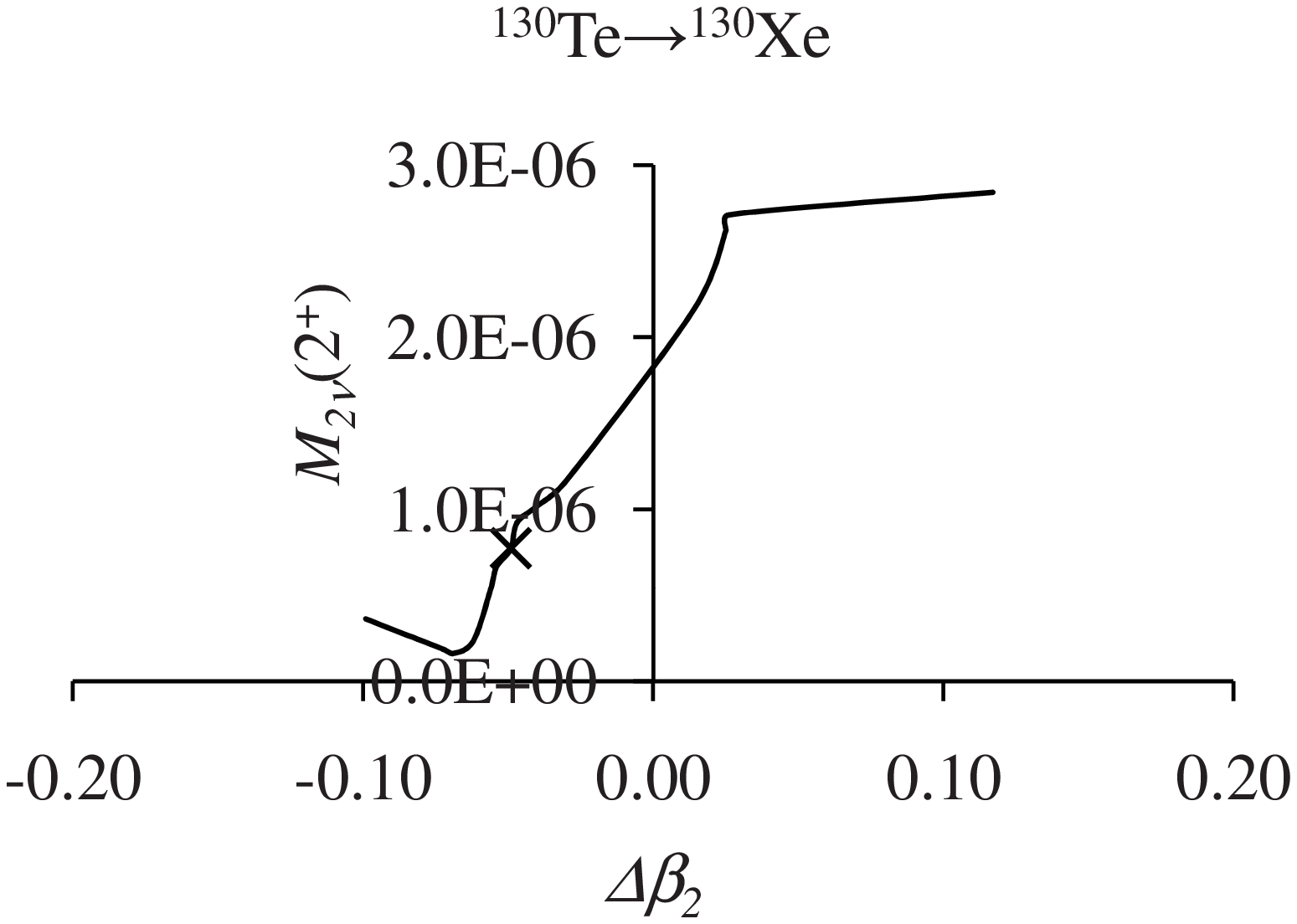}} &
\resizebox{0.45\textwidth}{!}{\includegraphics{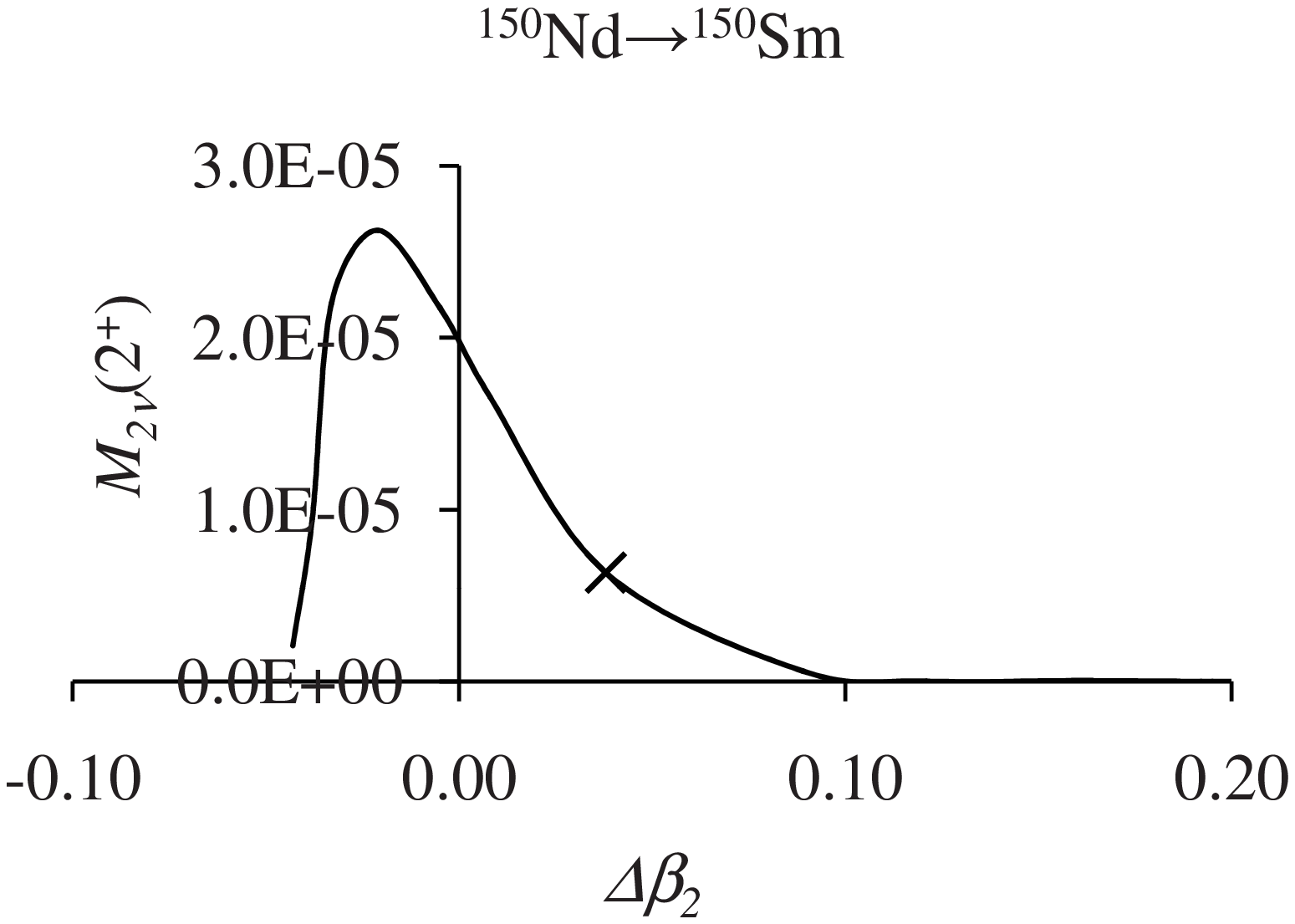}} \\
\end{tabular}
\caption{NTMEs as a function of the difference in the deformation parameter
$\Delta\beta_2$. \textquotedblleft $\times$ \textquotedblright\ denotes 
the value of calculated NTMEs for $\Delta\beta_2$ at $\zeta_{qq}=1.0$.}
\label{fig1}
\end{figure*}

As already mentioned, it has been observed that the inclusion of deformation
in the mean field can reduce the NTMEs $M_{2\nu }(2^{+})$ calculated in the
pnQRPA model up to a factor of 341 \cite{radu07}. In table 2, we present
the excitation energies $E_{2^{+}}$, quadrupole moments $Q(2^{+})$ of
daughter nuclei along, Q-values of $0^{+}\rightarrow 2^{+}$ transition $%
Q_{2^{+}}$ and the $G_{2\nu }(2^{+})$. According to the Grodzin's rule \cite%
{grod62}, the excitation energies $E_{2^{+}}$ and quadrupole moments $%
Q(2^{+})$ are inversely related. Although, a smaller $E_{2^{+}}$ can give a
higher Q-value $Q_{2^{+}}$ resulting in a larger phase space factor, the
NTMEs $M_{2\nu }(2^{+})$ are reduced due to a larger $Q(2^{+})$. Thus, the $%
0^{+}\rightarrow 2^{+}$ transition is intrinsically suppressed due to the
nuclear structure effects in addition to the cubic dependence of the energy
denominator.
\begin{table*}[htbp]
\caption{Excitation energies $E_{2^{+}}$, quadrupole moments $Q(2^{+})$ 
of daughter nuclei, Q-values of 0$^{+}\rightarrow $2$^{+}$
transition $Q_{2^{+}}$ and the phase space factors 
$G_{2\nu}(2^{+})$ with $g_{A}=1.2701$.}
\label{tab2}
\begin{tabular}{cllcccccc}
\hline\noalign{\smallskip}
{\small Transition} &~~~~~~~~~~~~~~~  & $E_{2^{+}}${\small \ (MeV)\cite{saka84}} 
&~~~~  & $Q${\small \ ($2^{+}$}){\small (eb)}{\small \cite{ragh89}} 
&~~~~~~~~~~  & $Q_{2^{+}}${\small \ (MeV)} &~~~~~~~~~~~  & ${\small G}_{2\nu }
{\small (2}^{+}{\small )}$ \\
\noalign{\smallskip}\hline\noalign{\smallskip}
$^{94}${\small Zr}$\rightarrow ^{94}${\small Mo} &  & {\small 0.871099} &  & 
{\small -0.13}$\pm ${\small 0.08} &  & {\small 1.145} &  & {\small 6.801}$%
\times 10^{-30}$ \\ 
$^{96}${\small Zr}$\rightarrow ^{96}${\small Mo} &  & {\small 0.778213} &  & 
{\small -0.20}$\pm ${\small 0.08} &  & {\small 2.572} &  & {\small 1.494}$%
\times 10^{-18}$ \\ 
\multicolumn{1}{c}{$^{100}${\small Mo}$\rightarrow ^{100}${\small Ru}} &  & 
{\small 0.53959} &  & {\small -0.54}$\pm ${\small \ 0.07} &  & {\small 2.494}
&  & {\small 1.460}$\times 10^{-18}$ \\ 
$^{104}${\small Ru}$\rightarrow ^{104}${\small Pd} &  & {\small 0.55579} & 
& {\small -0.47}$\pm ${\small \ 0.10} &  & {\small 0.743} &  & {\small 9.625}%
$\times 10^{-25}$ \\ 
\multicolumn{1}{c}{$^{110}${\small Pd}$\rightarrow ^{110}${\small Cd}} &  & 
{\small 0.657751} &  & {\small -0.40 }$\pm ${\small 0.04} &  & {\small 1.360}
&  & {\small 1.228}$\times 10^{-20}$ \\ 
\multicolumn{1}{c}{$^{128}${\small Te}$\rightarrow ^{128}${\small Xe}} &  & 
{\small 0.4429} &  &  &  & {\small 0.425} &  & {\small 1.429}$\times
10^{-24} $ \\ 
\multicolumn{1}{c}{$^{130}${\small Te}$\rightarrow ^{130}${\small Xe}} &  & 
{\small 0.5361} &  &  &  & {\small 1.989} &  & {\small 4.632}$\times
10^{-19} $ \\ 
\multicolumn{1}{c}{$^{150}${\small Nd}$\rightarrow ^{150}${\small Sm}} &  & 
{\small 0.33395} &  & {\small -1.32}$\pm ${\small 0.19} &  & {\small 3.037}
&  & {\small 3.253}$\times 10^{-17}$ \\
\noalign{\smallskip}\hline
\end{tabular}
\end{table*}

A large number of experimental and theoretical studies have been carried out
for the $0^{+}\rightarrow 2^{+}$ transition of \ $2\nu \beta ^{-}\beta ^{-}$
decay. Over the past years, the $0^{+}\rightarrow 2^{+}$ transition of $2\nu
\beta ^{-}\beta ^{-}$ decay\ of $^{94}$Zr \cite{norm87,doka16}, $^{96}$Zr 
\cite{bara96,arpe94}, $^{100}$Mo \cite{kudo92,blum92,bara93,bara95}, $^{110}$%
Pd \cite{lehn16}, $^{128}$Te \cite{bell87}, $^{130}$Te \cite{bell87,bara01}
and $^{150}$Nd \cite{arpe94,arpe99} isotopes has been experimentally
investigated. However, the $2\nu \beta ^{-}\beta ^{-}$ decay\ of $^{104}$Ru
for the 0$^{+}\rightarrow $2$^{+}$ transition has not been experimentally
investigated so far. All the available theoretical and experimental results
are compiled in table 3. We present only the theoretical $T_{1/2}^{2\nu }(2^{+})$
for those models for which no direct or indirect information about $M_{2\nu
}(2^{+})$ is available to us. As already mentioned, there is a remarkable
spread in the calculated NTMEs $M_{2\nu }(2^{+})$ within different models.
Specifically, the NTMEs $M_{2\nu }(2^{+})$ calculated with the QRPA model
without and with deformation vary by a factor of 2--341, corresponding to $%
^{130}$Te and $^{96}$Zr isotopes,respectively. The average NTMEs $\overline{M%
}_{2\nu }(2^{+})$ evaluated using the PHFB approach are suppressed by a
factor between 1\ -- 150 with respect to those of Raduta $et$ $al$. \cite%
{radu07} corresponding to $^{96}$Zr and $^{128}$Te isotopes, respectively.
Consideration of the available theoretical and experimental results suggests
that the prospective nuclei for the observation of the 0$^{+}\rightarrow $2$%
^{+}$ transition of $2\nu \beta ^{-}\beta ^{-}$ decay are $^{96}$Zr, $^{100}$%
Mo, $^{110}$Pd, $^{130}$Te and $^{150}$Nd.

\begin{table*}[htbp]
\caption{Theoretically calculated NTMEs $M_{2\nu }(2^{+})$ and
half-lives $T_{1/2}^{2\nu }(2^{+})$ for the $0^{+}\rightarrow 2^{+}$
transition of $^{94,96}$Zr, $^{100}$Mo, $^{104}$Ru, $^{110}$Pd, $^{128,130}$%
Te and $^{150}$Nd nuclei along with experimental half-lives $T_{1/2}^{2\nu
}(2^{+})$. \textquotedblleft *\textquotedblright\ denotes the present
calculation with the average NTME.}
\label{tab3}
\begin{tabular}{rlcllllclllll}
\hline\noalign{\smallskip}
{\small Nuclei} &  & \multicolumn{7}{c}{\small Theory} &  & 
\multicolumn{3}{c}{\small Experiment} \\ \cline{3-9}\cline{11-13}
&~~~~~~~~  & {\small Model} &~~  & {\small Ref.} &~~~~~~~  & $\left\vert M_{2\nu
}(2^{+})\right\vert $ &~~~~~~~  & $T_{1/2}^{2\nu }${\small (y)} &~~~~~~~  & 
$T_{1/2}^{2\nu}${\small (y)} &~~  & {\small Ref.} \\
\noalign{\smallskip}\hline\noalign{\smallskip}
\multicolumn{1}{r}{$^{94}${\small Zr}} &  & \multicolumn{1}{l}{\small PHFB}
&  & {\small *} &  & {\small 9.88}$\times ${\small 10}$^{-5}$ &  & {\small %
1.505}$\times ${\small 10}$^{37}$ &  & {\small{$>$}1.3}$\times $%
{\small \ 10}$^{19}$ &  & {\small \cite{norm87}} \\ 
\multicolumn{1}{l}{} &  & \multicolumn{1}{l}{{\small QRPA}$^{\dagger }$} & 
& {\small \cite{suho11}} &  & {\small 0.0170} &  & {\small 5.088}$\times $%
{\small 10}$^{32}$ &  & {\small {$>$}}3.4$\times ${\small \ 10}$%
^{19}$ &  & {\small \cite{doka16}} \\ 
\multicolumn{1}{l}{} &  & \multicolumn{1}{l}{{\small QRPA}$^{\ddagger }$} & 
& {\small \cite{suho11}} &  & {\small 0.0155} &  & {\small 6.120}$\times $%
{\small 10}$^{32}$ &  &  &  &  \\ 
\multicolumn{1}{r}{$^{96}${\small Zr}} &  & \multicolumn{1}{l}{\small PHFB}
&  & {\small *} &  & {\small 9.98}$\times ${\small 10}$^{-5}$ & 
\multicolumn{1}{l}{} & {\small 6.723}$\times ${\small 10}$^{25}$ &  & 
{\small {$>$}2.0}$\times ${\small \ 10}$^{18}$ &  & {\small \cite%
{norm87}} \\ 
\multicolumn{1}{l}{} &  & \multicolumn{1}{l}{\small QRPA} &  & {\small \cite%
{bara96}} &  & {\small (0.005-0.038)} & \multicolumn{1}{l}{} & {\small 2.677}%
$\times ${\small 10}$^{22}${\small -} &  & {\small {$>$}4.1}$\times 
${\small \ 10}$^{19}$ &  & {\small \cite{arpe94}} \\ 
\multicolumn{1}{l}{} &  & \multicolumn{1}{l}{} &  &  &  &  & 
\multicolumn{1}{l}{} & {\small 4.635}$\times ${\small 10}$^{20}$ &  & 
{\small {$>$}7.9}$\times ${\small \ 10}$^{19}$ &  & {\small \cite%
{bara96}} \\ 
\multicolumn{1}{l}{} &  & \multicolumn{1}{l}{\small QRPA} &  & {\small \cite%
{radu07}} &  & {\small 1.113}$\times ${\small 10}$^{-4}$ & 
\multicolumn{1}{l}{} & {\small 5.403}$\times ${\small 10}$^{25}$ &  &  &  & 
\\ 
\multicolumn{1}{l}{} &  & \multicolumn{1}{l}{\small QRPA} &  & {\small \cite%
{unlu14}} &  & {\small 0.011} & \multicolumn{1}{l}{} & {\small 5.532}$\times 
${\small 10}$^{21}$ &  &  &  &  \\ 
\multicolumn{1}{l}{} &  & \multicolumn{1}{l}{{\small RQRPA}$^{\dagger }$} & 
& {\small \cite{toiv97}} &  & {\small 0.011} & \multicolumn{1}{l}{} & 
{\small 5.532}$\times ${\small 10}$^{21}$ &  &  &  &  \\ 
\multicolumn{1}{l}{} &  & \multicolumn{1}{l}{{\small RQRPA}$^{\ddagger }$} & 
& {\small \cite{toiv97}} &  & {\small 0.010} & \multicolumn{1}{l}{} & 
{\small 6.693}$\times ${\small 10}$^{21}$ &  &  &  &  \\ 
\multicolumn{1}{l}{} &  & \multicolumn{1}{l}{\small RQRPA} &  & {\small \cite%
{schw98}} &  &  & \multicolumn{1}{l}{} & {\small (1.1-1.4)}$\times ${\small %
10}$^{21}$ &  &  &  &  \\ 
\multicolumn{1}{l}{} &  & \multicolumn{1}{l}{\small SRPA} &  & {\small \cite%
{stoi96}} &  & {\small 3.117}$\times ${\small 10}$^{-4}$ & 
\multicolumn{1}{l}{} & {\small 6.889}$\times ${\small 10}$^{24}$ &  &  &  & 
\\ 
\multicolumn{1}{r}{$^{100}${\small Mo}} &  & \multicolumn{1}{l}{\small PHFB}
&  & {\small *} &  & {\small 1.89}$\times ${\small 10}$^{-5}$ & 
\multicolumn{1}{l}{} & {\small 1.924}$\times ${\small 10}$^{27}$ &  & 
{\small {$>$}1.5}$\times ${\small \ 10}$^{20}$ &  & {\small \cite%
{kudo92}} \\ 
\multicolumn{1}{l}{} &  & \multicolumn{1}{l}{\small QRPA} &  & {\small \cite%
{suho94}} &  & {\small 0.033} & \multicolumn{1}{l}{} & {\small 6.290}$\times 
${\small 10}$^{20}$ &  & {\small {$>$}5.0}$\times ${\small \ 10}$%
^{20}$ &  & {\small \cite{blum92}} \\ 
\multicolumn{1}{l}{} &  & \multicolumn{1}{l}{\small QRPA} &  & {\small \cite%
{radu07}} &  & {\small 1.814}$\times ${\small 10}$^{-4}$ & 
\multicolumn{1}{l}{} & {\small 2.081}$\times ${\small 10}$^{25}$ &  & 
{\small {$>$}2.3}$\times ${\small \ 10}$^{21}$ &  & {\small \cite%
{bara93}} \\ 
\multicolumn{1}{l}{} &  & \multicolumn{1}{l}{\small QRPA} &  & {\small \cite%
{unlu14}} &  & {\small 0.0078} & \multicolumn{1}{l}{} & {\small 1.126}$%
\times ${\small 10}$^{22}$ &  & {\small {$>$}1.6}$\times ${\small \
10}$^{21}$ &  & {\small \cite{bara95}} \\ 
\multicolumn{1}{l}{} &  & \multicolumn{1}{l}{\small RQRPA} &  & {\small \cite%
{schw98}} &  &  & \multicolumn{1}{l}{} & {\small (1.0-1.1)}$\times ${\small %
10}$^{22}$ &  &  &  &  \\ 
\multicolumn{1}{l}{} &  & \multicolumn{1}{l}{\small SRPA} &  & {\small \cite%
{stoi96}} &  & {\small 1.482}$\times ${\small 10}$^{-3}$ & 
\multicolumn{1}{l}{} & {\small 3.119}$\times ${\small 10}$^{23}$ &  &  &  & 
\\ 
\multicolumn{1}{l}{} &  & \multicolumn{1}{l}{{\small SU(3)}$^{+}$} &  & 
{\small \cite{hirs95}} &  & {\small 7.3}$\times ${\small 10}$^{-5}$ & 
\multicolumn{1}{l}{} & {\small 1.285}$\times ${\small 10}$^{26}$ &  &  &  & 
\\ 
\multicolumn{1}{l}{} &  & \multicolumn{1}{l}{{\small SU(3)}$^{++}$} &  & 
{\small \cite{hirs95}} &  & {\small 1.53}$\times ${\small 10}$^{-4}$ & 
\multicolumn{1}{l}{} & {\small 2.926}$\times ${\small 10}$^{25}$ &  &  &  & 
\\ 
\multicolumn{1}{l}{} &  & \multicolumn{1}{l}{\small MCM} &  & {\small \cite%
{suho98b}} &  &  & \multicolumn{1}{l}{} & {\small (5.3-13)}$\times ${\small %
10}$^{20}$ &  &  &  &  \\ 
\multicolumn{1}{r}{$^{104}${\small Ru}} &  & \multicolumn{1}{l}{\small PHFB}
&  & {\small *} &  & {\small 3.63}$\times ${\small 10}$^{-5}$ &  & {\small %
7.867}$\times ${\small 10}$^{32}$ &  &  &  &  \\ 
\multicolumn{1}{l}{} &  & \multicolumn{1}{l}{\small QRPA} &  & {\small \cite%
{radu07}} &  & {\small 3.736}$\times ${\small 10}$^{-3}$ &  & {\small 7.444}$%
\times ${\small 10}$^{28}$ &  &  &  &  \\ 
\multicolumn{1}{l}{} &  & \multicolumn{1}{l}{{\small QRPA}$^{\dagger }$} & 
& {\small \cite{suho11}} &  & {\small 0.00792} &  & {\small 1.656}$\times $%
{\small 10}$^{28}$ &  &  &  &  \\ 
\multicolumn{1}{l}{} &  & \multicolumn{1}{l}{{\small QRPA}$^{\ddagger }$} & 
& {\small \cite{suho11}} &  & {\small 0.00811} &  & {\small 1.580}$\times $%
{\small 10}$^{28}$ &  &  &  &  \\ 
\multicolumn{1}{r}{$^{110}${\small Pd}} &  & \multicolumn{1}{l}{\small PHFB}
&  & {\small *} &  & {\small 1.19}$\times ${\small 10}$^{-4}$ &  & {\small %
5.731}$\times ${\small 10}$^{27}$ &  & {\small {$>$}2.9}$\times $%
{\small 10}$^{20}$ &  & {\small \cite{lehn16}} \\ 
\multicolumn{1}{l}{} &  & \multicolumn{1}{l}{\small QRPA} &  & {\small \cite%
{radu07}} &  & {\small 6.671}$\times ${\small 10}$^{-3}$ &  & {\small 1.830}$%
\times ${\small 10}$^{24}$ &  &  &  &  \\ 
\multicolumn{1}{l}{} &  & \multicolumn{1}{l}{{\small QRPA}$^{\dagger }$} & 
& {\small \cite{suho11}} &  & {\small 0.0112} &  & {\small 6.492}$\times $%
{\small 10}$^{23}$ &  &  &  &  \\ 
\multicolumn{1}{l}{} &  & \multicolumn{1}{l}{{\small QRPA}$^{\ddagger }$} & 
& {\small \cite{suho11}} &  & {\small 0.00766} &  & {\small 1.388}$\times $%
{\small 10}$^{24}$ &  &  &  &  \\ 
\multicolumn{1}{l}{} &  & \multicolumn{1}{l}{\small SRPA} &  & {\small \cite%
{stoi94}} &  & {\small 5.621}$\times ${\small 10}$^{-3}$ & 
\multicolumn{1}{l}{} & {\small 2.577}$\times ${\small 10}$^{24}$ &  &  &  & 
\\ 
\multicolumn{1}{r}{$^{128}${\small Te}} &  & \multicolumn{1}{l}{\small PHFB}
&  & {\small *} &  & {\small 2.07}$\times ${\small 10}$^{-6}$ & 
\multicolumn{1}{l}{} & {\small 1.636}$\times ${\small 10}$^{35}$ &  & 
{\small {$>$}4.7}$\times ${\small \ 10}$^{21}$ &  & {\small \cite%
{bell87}} \\ 
\multicolumn{1}{l}{} &  & \multicolumn{1}{l}{\small QRPA} &  & {\small \cite%
{radu07}} &  & {\small 3.055}$\times ${\small 10}$^{-4}$ & 
\multicolumn{1}{l}{} & {\small 7.498}$\times ${\small 10}$^{30}$ &  &  &  & 
\\ 
\multicolumn{1}{l}{} &  & \multicolumn{1}{l}{\small QRPA} &  & {\small \cite%
{unlu14}} &  & {\small 0.00287} & \multicolumn{1}{l}{} & {\small 8.496}$%
\times ${\small 10}$^{28}$ &  &  &  &  \\ 
\multicolumn{1}{l}{} &  & \multicolumn{1}{l}{\small SRPA} &  & {\small \cite%
{stoi94}} &  & {\small 1.022}$\times ${\small 10}$^{-3}$ & 
\multicolumn{1}{l}{} & {\small 6.700}$\times ${\small 10}$^{29}$ &  &  &  & 
\\ 
\multicolumn{1}{r}{$^{130}${\small Te}} &  & \multicolumn{1}{l}{\small PHFB}
&  & {\small *} &  & {\small 1.34}$\times ${\small 10}$^{-6}$ & 
\multicolumn{1}{l}{} & {\small 1.201}$\times ${\small 10}$^{30}$ &  & 
{\small {$>$}4.5}$\times ${\small \ 10}$^{21}$ &  & {\small \cite%
{bell87}} \\ 
\multicolumn{1}{l}{} &  & \multicolumn{1}{l}{\small QRPA} &  & {\small \cite%
{radu07}} &  & {\small 8.272}$\times ${\small 10}$^{-5}$ & 
\multicolumn{1}{l}{} & {\small 3.155}$\times ${\small 10}$^{26}$ &  & 
{\small {$>$}1.6}$\times ${\small \ 10}$^{21}$ &  & {\small \cite%
{bara01}} \\ 
\multicolumn{1}{l}{} &  & \multicolumn{1}{l}{\small QRPA} &  & {\small \cite%
{unlu14}} &  & {\small 0.00016} & \multicolumn{1}{l}{} & {\small 8.433}$%
\times ${\small 10}$^{25}$ &  &  &  &  \\ 
\multicolumn{1}{l}{} &  & \multicolumn{1}{l}{\small SRPA} &  & {\small \cite%
{stoi94}} &  & {\small 4.088}$\times ${\small 10}$^{-3}$ & 
\multicolumn{1}{l}{} & {\small 1.292}$\times ${\small 10}$^{23}$ &  &  &  & 
\\ 
$^{150}${\small Nd} &  & \multicolumn{1}{l}{\small PHFB} &  & {\small *} & 
& {\small 5.86}$\times ${\small 10}$^{-6}$ &  & {\small 8.940}$\times $%
{\small 10}$^{26}$ &  & {\small {$>$}8.0}$\times ${\small \ 10}$%
^{18}$ &  & {\small \cite{arpe94}} \\ 
&  & \multicolumn{1}{l}{\small SU(3)} &  & {\small \cite{hirs95b}} &  & 
{\small 5.38}$\times ${\small 10}$^{-5}$ &  & {\small 1.062}$\times 
${\small 10}$^{25}$ &  & {\small 
{$>$}9.1}$\times ${\small \ 10}$^{19}$ &  & {\small \cite{arpe99}}
\\
\noalign{\smallskip}\hline
\end{tabular}
\newline \noindent $^{\dagger }$WS basis; $^{\ddagger }$AWS basis; 
$^{+}$Spherical occupation wave functions; $^{++}$Deformed occupation 
wave functions
\end{table*}

\section{Conclusions}

Using a set of reliable wave functions generated with four different
parametrizations of the effective two-body interaction namely, \textit{PQQ}%
1, \textit{PQQHH}1, \textit{PQQ}2 and \textit{PQQHH}2 \cite%
{chan05,sing07,chan09,rath10}, sets of four NTMEs $M_{2\nu }(2^{+})$ have
been calculated to study the $2\nu \beta ^{-}\beta ^{-}$ decay of $^{94,96}$%
Zr, $^{100}$Mo, $^{104}$Ru, $^{110}$Pd, $^{128,130}$Te and $^{150}$Nd isotopes
for the $0^{+}\rightarrow 2^{+}$ transition. It is noticed that the $%
0^{+}\rightarrow 2^{+}$ transition is intrinsically suppressed due to the
cubic dependence of the energy denominator and nuclear structure effects.
Specifically, a large phase space factor due a larger Q-value implies a
smaller $E_{2^{+}}$ resulting from a larger $Q(2^{+})$, which results in the
suppression of NTMEs $M_{2\nu }(2^{+})$.

The observation of Raduta $et$ $al$. \cite{radu07} that the inclusion of
deformation in the mean field can reduce the NTMEs $M_{2\nu }(2^{+})$
calculated within pnQRPA up to a factor of 341, motivated us to study the 
$0^{+}\rightarrow 2^{+}$ transition of $2\nu \beta ^{-}\beta ^{-}$ decay
within PHFB approach treating the pairing and deformation degrees of freedom
simultaneously on equal footing. It is noticed that with respect to NTMEs $%
M_{2\nu }(2^{+})$ of Raduta $et$ $al$. \cite{radu07}, the average NTMEs $%
\overline{M}_{2\nu }(2^{+})$ calculated using the PHFB approach are further
suppressed by a factor between 1\ -- 150 corresponding to $^{96}$Zr and $%
^{128}$Te isotopes, respectively. In spite of the fact that the $%
0^{+}\rightarrow 2^{+}$ transition of $2\nu \beta ^{-}\beta ^{-}$ decay is
highly suppressed in comparison to the $0^{+}\rightarrow 0^{+}$ transition,
the available theoretical and experimental results suggest that the
observation of the $0^{+}\rightarrow 2^{+}$ transition of $2\nu \beta
^{-}\beta ^{-}$ decay may be possible in $^{96}$Zr, $^{100}$Mo, $^{110}$Pd, $%
^{130}$Te and $^{150}$Nd isotopes.

\acknowledgement {This work is partially supported by DST-SERB, India
vide sanction No. SR/FTP/PS-085/2011, SB/S2/HEP-007/2013 and Council of
Scientific and Industrial Research (CSIR), India vide sanction No.
03(1216)/12/EMR-II}.

\end{document}